\documentclass[times, review, 10pt]{elsarticle}

\usepackage[top=4.3cm,right=4.8cm,bottom=4.3cm,left=4.8cm]{geometry}
\usepackage{setspace}
\usepackage{amssymb}
\usepackage{array}   
\usepackage{amsmath}

\newcolumntype{C}[1]{>{\centering\arraybackslash}p{#1}}
\usepackage{amsmath}

\usepackage{subcaption}
\usepackage{url}
\usepackage{booktabs}
\usepackage{adjustbox}
\usepackage{tabularx}
\usepackage{makecell}
\usepackage{xurl} 
\usepackage{graphicx}
\usepackage{multirow}
\usepackage{float}
\usepackage[table]{xcolor}
\usepackage{colortbl}

\begin{document}

\begin{frontmatter}

\title{Robust Audio Tagging under Class-wise Supervision Unreliability}

\author[ox]{Yuanbo Hou\corref{cor1}}
\author[lu]{Zhaoyi Liu\fnref{eq}}
\author[gis]{Tong Ye\fnref{eq}}
\author[kth]{Qiaoqiao Ren}
\author[gis]{Jian Guan}
\author[surry]{Wenwu Wang}
\author[ox]{Stephen Roberts}

\affiliation[ox]{organization={Machine Learning Research Group, Engineering Science, University of Oxford}, country={UK}}
\affiliation[lu]{organization={KU Leuven}, country={Belgium}} 
\affiliation[gis]{organization={GISP, Harbin Engineering University}, country={China}}
\affiliation[kth]{organization={EECS, KTH Royal Institute of Technology}, country={Sweden}}
\affiliation[surry]{organization={CVSSP, University of Surrey}, country={UK}}

\fntext[eq]{Equal contribution.}
\cortext[cor1]{Corresponding author: Yuanbo Hou, Machine Learning Research Group, University of Oxford, UK. Email: Yuanbo.Hou@eng.ox.ac.uk}

\begin{abstract}  
Weakly labeled datasets such as AudioSet have driven recent progress in audio tagging. However, annotation quality varies across sound classes. Labels may be incomplete, ambiguous, or unreliable, which introduces class-dependent supervision bias during optimisation. The issue becomes harder as real and generated audio are increasingly mixed in training, and generated samples do not always match their intended semantic labels. Prior work mainly addressed unreliable supervision from missing-positive labels, while this paper targets three other sources of unreliable supervision: spurious additions, misassignments between similar classes, and weakened label evidence. These effects introduce class-dependent optimisation bias that is not explicitly modeled by most existing methods. To bridge this gap, the paper proposes a Class-wise Supervision Unreliability (CSU) framework that controls supervision strength at the class level during training. CSU learns a separate unreliability parameter for each class and down-weights less reliable supervision without changing the model architecture or inference process. To support evaluations, this paper also introduces ESC-FreeGen50, a manually verified benchmark of 50 sound classes that combines real and generated audio. Experiments on controlled benchmarks and AudioSet show that CSU improves robustness across different architectures and different sources of supervision unreliability. The results indicate that explicit class-wise modeling of supervision unreliability is an effective and practical strategy for robust audio tagging under large-scale weakly labeled training. Code and data are available at: \textcolor{blue}{\underline{https://github.com/Yuanbo2020/CSU}}. 
\end{abstract}


 
\end{frontmatter}
 
\section{Introduction}
\label{Introduction}  
Recent progress in audio event classification (AEC) and audio tagging (AT) has been driven by large-scale weakly labeled datasets such as AudioSet \cite{AudioSet} and FSD50K \cite{Fonseca2022fsd50k}. 
These datasets support transferable representation learning from large audio collections and have advanced environmental sound recognition, acoustic scene analysis, and soundscape captioning \cite{Soundscape}. 
Despite these gains, weakly labeled audio corpora still suffer from a basic limitation: supervision quality is not uniform across sound classes. Annotations collected from web metadata and non-expert curation are often incomplete, ambiguous, or unreliable, which can introduce systematic class-dependent bias during optimisation and reduce model robustness \cite{fonseca2019audio}.   

Existing work often targets unreliable supervision caused by missing labels, particularly missing-positive labels \cite{fonseca2020addressing, fonseca2019learning}, where present sound events are left unannotated. Approaches such as teacher-student learning and loss masking have shown strong effectiveness in mitigating this issue \cite{fonseca2020addressing}.  
However, missing-positive labels capture only part of the supervision unreliability observed in real-world audio data \cite{iqbal2022noisy} and are not the focus of this paper, as shown in Table \ref{tab:noise_types}. 
Building on empirical observations from large-scale and community-curated corpora \cite{AudioSet, Fonseca2022fsd50k}, the paper studies the following corruption types, which introduce systematic class-wise optimisation bias, as shown in Table \ref{tab:noise_types}:
1) Spurious addition noise (SAN): non-existent events are erroneously tagged as present, e.g., a \textit{tick-tock} label assigned to a clip dominated by \textit{human speech} \cite{AudioSet};
2) Misassignment noise (MAN): labels are assigned to the wrong but perceptually similar class, e.g., \textit{tick-tock} versus \textit{dripping water}, resulting in incorrect class attribution \cite{gong2021psla};
3) Soft label noise (SLN): genuine events are tagged as present, but label evidence is weakened and represented as a soft probability rather than a fully confident positive label, e.g., a \textit{rain} event masked by traffic may be assigned a target such as 0.6 instead of 1.0 \cite{mendez2022eliciting}.
Although these corruption types arise differently, they share the same consequence: class-dependent optimisation bias. The problem becomes more pronounced in polyphonic and temporally overlapping audio, where incorrect labels, mixed evidence, and class confusions are difficult to disentangle \cite{baelde2019real}. 
Therefore, a training mechanism is needed that can respond to class-wise supervision unreliability without assuming an identifiable instance-wise corruption path or a noise-type-specific correction rule.
  
\begin{table}[t]
\footnotesize
\centering
\setlength{\abovecaptionskip}{0.1cm}   
\setlength{\belowcaptionskip}{-0.4cm}  
\setlength{\tabcolsep}{4pt} 
\renewcommand{\arraystretch}{1.1}

\begin{tabularx}{\linewidth}{
    >{\raggedright\arraybackslash}p{1.5cm}
    >{\raggedright\arraybackslash}X
    >{\raggedright\arraybackslash}X
    >{\centering\arraybackslash}p{1.5cm}
}
\toprule
\textbf{Type} & \textbf{Name of label corruption} & \textbf{Clean $\rightarrow$ Corrupted} & \textbf{In scope?} \\
\midrule
Type-0   & Missing-positive label         & [0, 0, 1, 0] $\rightarrow$ [0, 0, 0, 0]   & No  \\
Type-I   & Spurious-addition noise (SAN) & [1, 0, 1, 0] $\rightarrow$ [1, 1, 0, 0]   & Yes \\
Type-II  & Misassignment noise (MAN)     & [0, 0, 1, 0] $\rightarrow$ [0, 1, 0, 0]   & Yes \\
Type-III & Soft-label noise (SLN)        & [0, 0, 0, 1] $\rightarrow$ [0, 0, 0, 0.6] & Yes \\
\bottomrule
\end{tabularx}
\caption{ 
    Examples of four label-corruption types; Type-0 has been well studied in prior work and is not considered in this paper. Assume the label order is [\textit{speech}, \textit{tick-tock}, \textit{dripping water}, \textit{rain}].
}
\label{tab:noise_types}
\end{table}  
 
Robust learning under noisy labels has been widely studied, but most existing methods do not directly address the above class-dependent supervision unreliability in weakly labeled audio tagging (AT). Recent studies improve noisy-label learning by expanding the problem setting or making supervision reliability more explicit. Open-set modeling \cite{zhang2025openset} considers supervision noise beyond closed-set assumptions. Progressive sample selection \cite{zhang2025psscl} instead tries to separate cleaner samples from corrupted ones during training. Uncertainty-based methods \cite{zhou2026affinity} estimate label reliability before using supervision for optimisation. Other approaches make training less sensitive to corrupted supervision. Symmetric Cross Entropy (SCE) \cite{wang2019sce} reduces the effect of noisy labels through a more robust loss design. Bootstrapping \cite{Bootstrapping} and uncertainty-aware pseudo-label selection (UPS) \cite{ups} revise targets with model predictions or confidence estimates. Asymmetric Loss (ASL) \cite{ridnik2021_asl} and Asymmetric Polynomial Loss (APL) \cite{huang2023apl} suppress the influence of likely mislabels during optimisation. Related ideas also appear under incomplete feedback, where $\rho$-corrected sequential Dynamic Classification \cite{lowne2010sequential} handles non-stationary supervision when feedback is intermittent. Despite their differences, these methods usually assume that unreliable supervision can be identified and corrected at the level of labels, targets, or updates for individual instances. Weakly labeled polyphonic audio tagging rarely provides that level of observability. Clip-level labels do not reveal which event is wrong, which class is confused, or which part of the mixture should be trusted. The problem becomes harder when real and generated audio are trained together. Under these conditions, class-wise supervision control is more realistic than instance-wise corruption modeling or noise-type-specific correction in large-scale weakly labeled AT.
 
Recent use of synthetic audio introduces an increasingly relevant source of supervision unreliability. Generative models such as AudioLDM2 \cite{liu2024audioldm} are now widely used to synthesize training samples, augment scarce classes, or balance long-tailed distributions \cite{ghosh2025synthio}; however, generated audio does not always faithfully represent the intended semantic labels. Synthetic clips may contain subtle artifacts, mixed-source characteristics, or ambiguous acoustic cues, which reduce label consistency and further degrade supervision reliability \cite{huang2023make}. In practice, such generative ambiguity manifests as supervision unreliability, typically as weak or mixed evidence and occasional spurious cues, which aligns with SLN and SAN. As real and generated recordings become increasingly intertwined in data pipelines, label noise and generative artefacts can compound each other \cite{muller2022human}, producing stronger and more heterogeneous class-dependent supervision unreliability, not just a larger amount of generic label noise. 

These observations motivate a shift from repairing individual corrupted labels to regulating optimisation under class-dependent supervision unreliability. To address this problem, the proposed Class-wise Supervision Unreliability (CSU) framework models supervision unreliability as a class-level property. Each sound class is assigned a learnable unreliability parameter $\boldsymbol{\sigma}$ that controls how strongly supervision from that class contributes to the training objective. Larger $\boldsymbol{\sigma}$ values indicate less reliable supervision and produce stronger down-weighting during optimisation. Instead of trying to identify which individual labels should be repaired, filtered, or relabeled, CSU learns how strongly supervision from each class should influence optimisation. Therefore, CSU acts as a class-wise supervision control mechanism for SAN, MAN, and SLN, while remaining architecture-agnostic and requiring no inference modification.  
  
The evaluation setting should reflect the same problem. If the aim is to test robustness to class-dependent supervision unreliability, the benchmark must separate that factor from other sources of variation in weakly labeled audio corpora. Large-scale corpora, such as AudioSet \cite{AudioSet}, do not provide manually verified labels and controlled conditions that isolate SAN, MAN, and SLN. Weakly labeled clip-level annotations in real polyphonic audio also do not support identifiable instance-level corruption modeling \cite{AudioSet}. Hence, this paper introduces ESC-FreeGen50, a manually verified hybrid benchmark that combines real and generated audio and provides controlled corruption settings for SAN, MAN, and SLN. ESC-FreeGen50 and AudioSet serve complementary roles: the former isolates the target supervision unreliability under controlled conditions, and the latter tests whether the resulting gains remain relevant at scale.
 
Our contributions are threefold: 1) a class-wise supervision control framework, CSU, for weakly labeled audio tagging under class-dependent supervision unreliability, providing a unified training strategy for SAN, MAN, and SLN without inference modification; 2) ESC-FreeGen50, a manually verified real-generated benchmark with controlled corruption settings for SAN, MAN, and SLN, designed to make class-dependent supervision unreliability directly measurable beyond what large-scale weakly labeled corpora alone can provide; and 3) extensive experiments on ESC-FreeGen50 and AudioSet showing that CSU improves robustness across corruption types and model architectures under both controlled and large-scale weakly labeled settings. 
The remainder of this paper is organised as follows.
Section \ref{section_data} introduces the ESC-FreeGen50 dataset.
Section \ref{section_clu} presents the CSU framework.
Section \ref{section_experiments} describes the experimental setup and analyses the results.
Section \ref{section_conclusion} concludes the paper.

\section{The hybrid benchmark: ESC-FreeGen50}\label{section_data}  
 
Generated audio has become a practical way to expand data for audio classification and tagging \cite{ghosh2025synthio}. As real and generated recordings are increasingly used in combination, studying supervision unreliability under controlled conditions has become more difficult. 
While existing small-scale, clean benchmark datasets like ESC-50 \cite{piczak2015esc} have reliable labels, they lack characteristics of mixed real-generated audio. Large-scale, weakly labeled corpora like AudioSet \cite{AudioSet} and FSD50K \cite{Fonseca2022fsd50k} better reflect real-world polyphonic audio, but they do not provide human-verified labels or controlled conditions for isolating SAN, MAN, and SLN.  
Motivated by this gap, ESC-FreeGen50 is introduced as a human-verified hybrid dataset for controlled evaluation under SAN, MAN, and SLN, while covering the same 50 sound-event classes as ESC-50 across real and generated recordings. 
Specifically, ESC-50 \cite{piczak2015esc} is extended using a curated selection of real-world recordings from Freesound \cite{Font2013FreesoundTD}, and further extended with audio clips generated by a text-to-audio framework.
This design maintains label consistency and better reflects the current state of data for AT tasks.
To ensure label reliability, all audio clips in ESC-FreeGen50 undergo manual review and cross-checking for semantic correctness, perceptual clarity, and category consistency.  
The dataset homepage is: 
\textcolor{blue}{\underline{https://github.com/Yuanbo2020/ESC-FreeGen50}} .

\subsection{Dataset Composition}
  
ESC-FreeGen50 contains 50\% real recordings and 50\% generated recordings.

\subsubsection{Real-world Recordings}
\textbf{ESC-50} \cite{piczak2015esc} is a widely used environmental sound classification dataset. It covers 50 semantic categories, each containing 40 audio clips, for a total of 2,000 5-second clips. In ESC-FreeGen50, ESC-50 provides the backbone taxonomy and the clean real-audio base. 

\textbf{Freesound} \cite{Font2013FreesoundTD} is used to extend the real-audio component. Freesound is a large, community-driven repository with diverse real-world sounds. To match the ESC-50 taxonomy, recordings are manually curated class by class. Each selected clip is trimmed to 5 seconds and screened to ensure one dominant sound event and clear class consistency. This process adds 10 real recordings per class and increases acoustic variability within the real-audio portion of the dataset. 

\begin{figure}[t]
\setlength{\abovecaptionskip}{0.1cm}   
	\setlength{\belowcaptionskip}{-0.1cm}  
    \centering
    \includegraphics[width=0.8\linewidth]{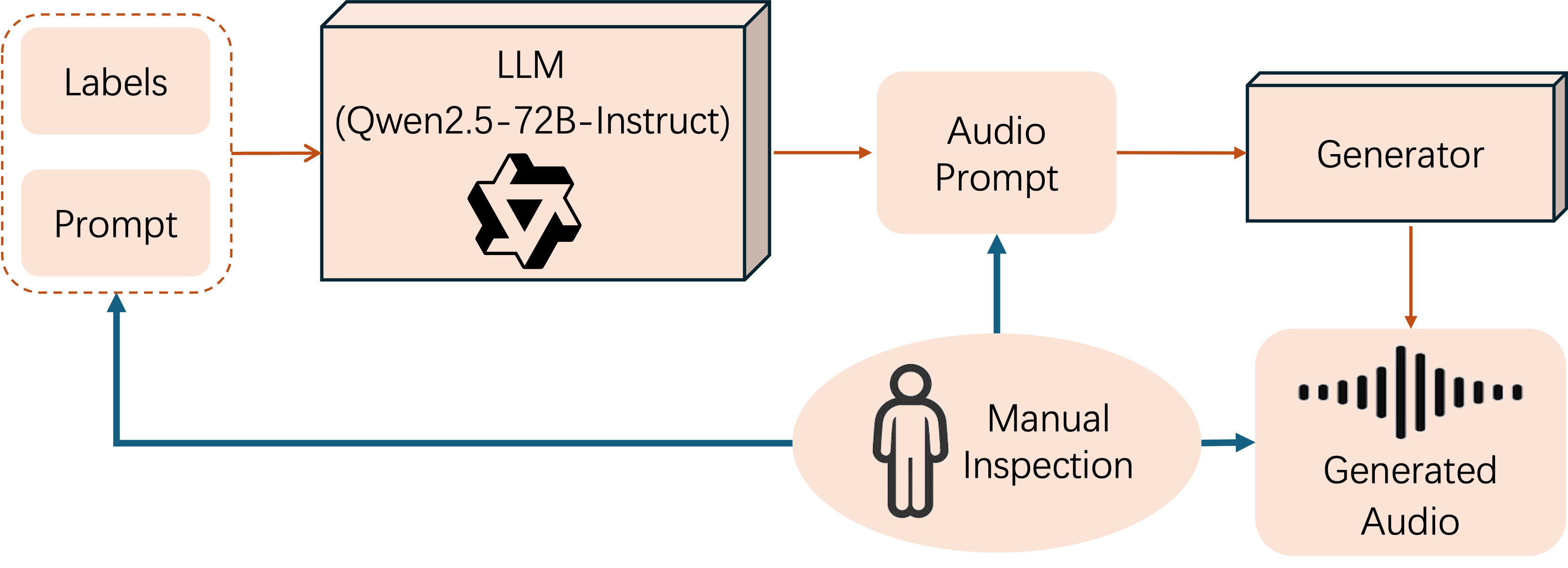}
    \caption{The data generation pipeline used to construct ESC-FreeGen50.}
    \label{fig:gen_audio}
\end{figure}

\subsubsection{Generated Recordings}
\begin{figure}[t]
\setlength{\abovecaptionskip}{0.1cm}   
	\setlength{\belowcaptionskip}{-0.1cm}  
    \centering
    \includegraphics[width=1\linewidth]{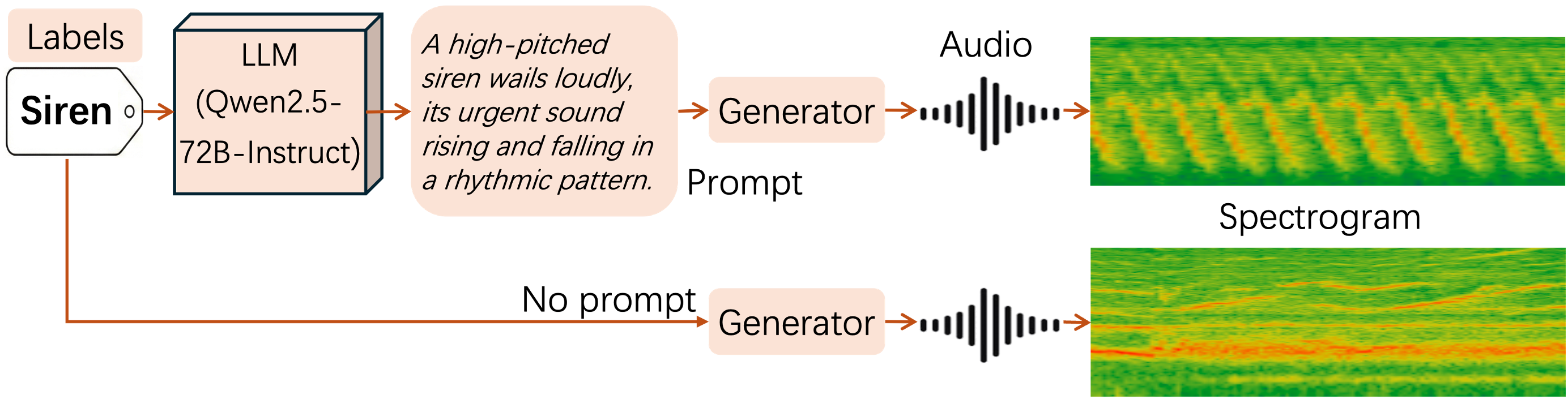}
    \caption{Using \textit{siren} as an example, the figure compares two strategies for audio generation. The LLM-based strategy expands the label into a descriptive prompt, where Qwen2.5-72B-Instruct is used for prompt expansion, with system prompt settings released on the homepage. AudioLDM2 \cite{liu2024audioldm} is used as the current generator. The LLM-based prompt yields a clearer rhythmic pattern in the spectrogram, which is more consistent with everyday \textit{siren} acoustics. Audio examples are available on the homepage.}
    \label{fig:prompt_comparison}
\end{figure}

Generated recordings are included because ESC-FreeGen50 is intended to reflect current AT settings in which mixed real and generated audio are used. As shown in Fig.~\ref{fig:gen_audio}, a text-to-audio generation pipeline is used to synthesise class-consistent audio for each ESC-50 category. 
The current implementation uses AudioLDM2 \cite{liu2024audioldm} as the generator, while keeping the overall framework open to future generator replacement. 
In practice, two strategies are explored during dataset construction: 
1) directly using the ESC-50 class label as the text prompt; 
2) using a Large Language Model (LLM) to expand each class label into a descriptive audio prompt, followed by manual review before generation.
During manual screening, LLM-derived prompts produced generated clips with clearer semantic content and stronger perceptual consistency than direct label prompts. Fig.~\ref{fig:prompt_comparison} shows a sample of these two strategies. Direct label-based generation often produces acoustically ambiguous or weakly instantiated sound events, whereas LLM-guided prompts provide richer contextual cues that better constrain the generation process. Accordingly, the LLM-based prompt strategy is adopted for all generated recordings in ESC-FreeGen50.
For each class, generated candidates are screened manually to remove semantically incorrect samples, poor-quality outputs, and clips with obvious artefacts. This process yields 50 validated generated recordings per class. For more details, please see the dataset homepage.
 
In total, ESC-FreeGen50 contains 100 clips per class and 5,000 clips overall. Of these, 40 clips per class come from ESC-50, 10 come from curated Freesound recordings, and 50 come from the generation pipeline. All clips are five seconds long, and all labels are manually verified.

\subsection{Role of ESC-FreeGen50 in This Study}
ESC-FreeGen50 is introduced as a clean and controlled benchmark for evaluating model robustness under SAN, MAN, and SLN.
Because its labels are manually verified, corruption can be injected from a reliable starting point, making it easier to separate the effects of the corruption itself. The balanced combination of real and generated audio also supports evaluation in mixed-source scenarios. ESC-FreeGen50 and AudioSet play different roles in the experiments. 
ESC-FreeGen50 is used for controlled analysis of SAN, MAN, and SLN; AudioSet is used for large-scale validation under real, weakly labeled polyphonic conditions. 

\vspace{-2mm}
\subsection{Dataset Partitioning}\label{partitioning}
ESC-FreeGen50 is divided into training, validation, and test subsets for model training, selection, and final evaluation. All splits are class-balanced and preserve the balance between real and generated data sources.
\textbf{Training set:} 4,000 samples (50 classes$\times$80 samples/class), used for model training.
\textbf{Validation set:} 500 samples (50 classes$\times$10 samples/class), used for hyperparameter tuning and early stopping.
\textbf{Test set:} 500 samples (50 classes$\times$10 samples/class), held out for final evaluation. All test samples are disjoint from those in the training and validation sets.

\vspace{-2mm}
\section{Class-wise Supervision Unreliability}\label{section_clu}
 
Under weak clip-level multi-label supervision, SAN, MAN, and SLN arise from different annotation mechanisms but share a common learning consequence: they distort optimisation in a class-dependent manner. SAN introduces spurious positive pressure on absent events. MAN creates contradictory supervision between confusable classes. SLN preserves class identity but weakens the effective evidence supporting the positive label. Under weak labels, these supervision effects are difficult to disentangle at the individual sound-event instance level, but their common class-level optimisation effects remain visible during training. This section focuses on class-level supervision control rather than instance-level corruption recovery.  

\subsection{Rationale for Class-wise Supervision Unreliability}
\label{section_Rationale}

Analyses of web-sourced audio corpora such as FSDnoisy18k \cite{fonseca2019learning} and FSDKaggle \cite{fonseca2019audio} show that supervision reliability can vary substantially across sound classes. In FSDnoisy18k, estimated per-class noise rates range from roughly 20\% to over 80\% \cite{fonseca2019learning}. 
Such variation suggests that supervision unreliability is not evenly distributed across classes. 
Consequently, prior noisy-label studies often model corruption as class-conditional rather than uniform across categories \cite{natarajan2013learning, song2022learning}. However, weakly labeled polyphonic audio datasets with clip-level annotations do not reveal how corruption affects individual sound-event instances within a clip. Real-world mixtures contain overlapping sounds, ambiguous evidence, and incomplete annotation. Under these conditions, a full class-conditional noise transition matrix is difficult to identify and estimate reliably in practice \cite{liu2023identifiability}. 

These observations explain why this paper discusses SAN, MAN, and SLN together. Their source mechanisms differ, but weak clip-level labels expose their shared optimisation consequence: some classes receive supervision that remains less reliable than others during training. Therefore, the practical modeling target is not the latent corruption path for each instance, but the class-wise tendency of supervision reliability during training. Existing robust learning methods \cite{ridnik2021_asl, huang2023apl} also suggest that unreliable supervision should contribute less during optimisation. These considerations motivate Class-wise Supervision Unreliability (CSU). In CSU, each sound event class is assigned a learnable positive scalar $\sigma_i>0$ that controls its effective supervision strength during training, where a larger $\sigma_i$ indicates higher supervision unreliability. The scalar is not introduced to recover hidden clean labels or to estimate explicit corruption transitions. Its role is to modulate how strongly each class influences optimisation when its supervision remains unreliable. In the formulation below, $\sigma_i$ serves as a class-wise supervision control variable that adjusts the contribution of class $i$ to the training objective and the resulting parameter updates. This class-wise supervision-control view is examined empirically in Section~\ref{section_experiments} through results under SAN, MAN, and SLN and the learned class-wise $\sigma$ patterns. 

\vspace{-2mm}
\subsection{Class-wise Supervision Unreliability Modeling}
 
Based on the rationale above, CSU is formulated as a class-wise supervision-control mechanism within a standard sigmoid-based multi-label setting. The aim is to regulate how strongly supervision from each class influences optimisation when reliability differs across classes. We formalise this idea in two steps: 1) an exact scaled-logit form is introduced to make the optimisation effect of class-wise supervision control explicit; 2) a practical surrogate objective is derived for stable training and alignment with the standard binary cross-entropy (BCE) formulation.

Let $X$ denote the input feature representation of an audio clip. A Neural Network (NN) with parameters $W$ produces a logit vector $f^W(X) = (f_1,\dots,f_C)\in\mathbb{R}^C$, where $C$ is the number of sound classes. 
The observed multi-label target is $Y=(y_1,\dots,y_C)$ with $y_i\in[0,1]$ indicating the target value for class $i$. 
Following standard sigmoid-based multi-label classification, the per-class labels are treated as conditionally independent given the network outputs and the class-wise supervision unreliability parameters $\boldsymbol\sigma$. The joint likelihood for a single audio clip is 
\begin{equation}
\setlength{\abovedisplayskip}{1pt}
\setlength{\belowdisplayskip}{1pt}
p\bigl(Y \mid f^W(X), \boldsymbol\sigma\bigr)
= \prod_{i=1}^{C} p\bigl(y_i \mid f_i,\sigma_i\bigr)
\label{joint-likelihood}
\end{equation}

\subsubsection{Bernoulli Likelihood with Class-wise Supervision Scaling} 

For each class $i$, a Bernoulli form is used.
To encode class-wise supervision unreliability, the contribution of each class is modulated through a positive scalar $\sigma_i>0$ that controls the effective supervision strength of class $i$ during training. Here, $\sigma_i$ is interpreted as a learnable class-wise supervision control variable, rather than as an explicit estimate of a noise transition probability or an instance-level corruption state. The scaled logit is defined as $z_i := f_i/\sigma_i^2$, with corresponding success probability $p_i$
\begin{equation}
\setlength{\abovedisplayskip}{1pt}
\setlength{\belowdisplayskip}{1pt} 
    p_i = \mathrm{Sigmoid}(z_i),
    \quad
    p\bigl(y_i \mid f_i,\sigma_i\bigr)
    = p_i^{\,y_i} (1-p_i)^{1-y_i}
    \label{bernoulli-clu}
\end{equation} 
Here, $p_i\in(0,1)$ denotes the model-predicted probability for class $i$ under class-wise supervision scaling. Larger values of $\sigma_i$ shrink the magnitude of $z_i$, making the prediction for class $i$ less confident and reducing the influence of less reliable supervision for that class. 
The scaled-logit form therefore controls how strongly each class contributes to optimisation.
The Negative Log-Likelihood (NLL) for class $i$ is
\begin{equation}
\setlength{\abovedisplayskip}{1pt}
\setlength{\belowdisplayskip}{1pt}
- \log p(y_i \mid f_i,\sigma_i)
= - y_i \log p_i - (1-y_i)\log(1-p_i)
\label{aleatory-exact-nll}
\end{equation}
Using the standard logistic identity, this becomes
\begin{equation}
\setlength{\abovedisplayskip}{1pt}
\setlength{\belowdisplayskip}{1pt}
- \log p(y_i \mid f_i,\sigma_i)
= \log\bigl(1+\exp(z_i)\bigr) - y_i z_i
\label{eq:nll-z}
\end{equation} 
Differentiating \eqref{eq:nll-z} with respect to the raw logit $f_i$ yields
\begin{equation}
\setlength{\abovedisplayskip}{1pt}
\setlength{\belowdisplayskip}{1pt}
\frac{\partial}{\partial f_i} \bigl[-\log p(y_i \mid f_i,\sigma_i)\bigr]
= \frac{1}{\sigma_i^2}\bigl(\mathrm{Sigmoid}(z_i)-y_i\bigr)
\label{eq:grad-attenuation}
\end{equation}
 
Equation \eqref{eq:grad-attenuation} makes the role of CSU explicit. A larger $\sigma_i$ introduces a $1/\sigma_i^2$ prefactor and simultaneously shrinks the scaled logit $z_i=f_i/\sigma_i^2$. As a result, supervision from class $i$ produces weaker optimisation updates and less confident class-\(i\) predictions. This also clarifies how CSU responds to the three corruption types: for SAN, CSU suppresses spurious positive reinforcement on absent events; for MAN, CSU reduces the optimisation damage caused by contradictory supervision among confusable classes; for SLN, CSU down-weights supervision whose class identity is preserved but whose positive evidence is weakened. In all three cases, the learned $\sigma_i$ attenuates the optimisation influence of supervision that remains persistently less reliable for class $i$.
   
Next, to align CSU with the standard BCE objective used in audio tagging, we introduce a practical surrogate objective.

\vspace{-2mm}
\subsubsection{Practical Surrogate Objective}
The NLL in \eqref{eq:nll-z} depends on $\sigma_i$ through the scaled logit $z_i=f_i/\sigma_i^2$ inside the nonlinear $\log(1+\exp(\cdot))$ term. Although the objective can be optimised directly, the coupling between $f_i$ and $\sigma_i$ makes the role of $\sigma_i$ less transparent relative to the standard BCE used in AT tasks. Thus, we introduce a practical surrogate objective that remains aligned with BCE training while preserving the intended attenuation behaviour of CSU. It is designed to satisfy three points: consistency with standard BCE when $\sigma_i=1$, decreasing supervision contribution as $\sigma_i$ increases, and regularisation of $\sigma_i$.

For a single audio sample, the BCE for class $i$ using the unscaled logit $f_i$ is
\begin{equation}
\setlength{\abovedisplayskip}{1pt}
\setlength{\belowdisplayskip}{1pt}
L_i(W) = \log\bigl(1+\exp(f_i)\bigr) - y_i f_i
\label{eq:standard-bce}
\end{equation}
The surrogate objective is then written as
\begin{equation}
\setlength{\abovedisplayskip}{1pt}
\setlength{\belowdisplayskip}{1pt}
\mathcal{L}_{\text{surr},i}(W,\sigma_i) := a(\sigma_i)L_i(W) + b(\sigma_i)
\label{eq:surrogate-ansatz}
\end{equation}
The scalar functions $a(\sigma_i)$ and $b(\sigma_i)$ are selected to satisfy the following properties.

\paragraph{Consistency with BCE}
When $\sigma_i = 1$, the surrogate should recover standard BCE training up to an additive constant, so that gradients with respect to $W$ remain unchanged. This requires $a(1)=1$, while $b(1)$ may be any constant.
 
\paragraph{Supervision Attenuation}
As $\sigma_i$ increases, the contribution of class $i$ to the training gradients should decrease to reflect higher supervision unreliability. To match the exact gradient prefactor in \eqref{eq:grad-attenuation}, the weighting term is chosen as $a(\sigma_i) = 1/{\sigma_i^2}$.
 
\paragraph{Regularisation of \(\sigma_i\)}
To avoid the trivial solution $\sigma_i \rightarrow \infty$, where the weighted loss vanishes, the surrogate includes a penalty term that grows monotonically with $\sigma_i$. A logarithmic regulariser is adopted: 
$b(\sigma_i) = \log(\sigma_i + 1)$, which grows sublinearly and provides a mild penalty for large $\sigma_i$. The regulariser discourages large $\sigma_i$, while the positive parameterisation ensures $\sigma_i>0$. Other regularisers are possible, but this choice provides a simple and effective instantiation.

Substituting these choices into \eqref{eq:surrogate-ansatz} yields the surrogate loss for class $i$
\begin{equation}
\setlength{\abovedisplayskip}{1pt}
\setlength{\belowdisplayskip}{1pt}
\mathcal{L}_{\text{surr},i}
= \frac{1}{\sigma_i^2} L_i(W) + \log(\sigma_i + 1)
\label{nll_approx}
\end{equation}
For the surrogate objective,
\begin{equation}
\setlength{\abovedisplayskip}{2pt}
\setlength{\belowdisplayskip}{2pt}
\frac{\partial \mathcal{L}_{\text{surr},i}}{\partial W}
=
\frac{1}{\sigma_i^2}\frac{\partial L_i(W)}{\partial W}
\end{equation}
so the contribution of class $i$ to the network-parameter update is explicitly down-weighted as $\sigma_i$ increases. This makes the effect of class-wise supervision control explicit during optimisation. When $\sigma_i=1$, the surrogate recovers standard BCE up to an additive constant, i.e., $\mathcal{L}_{\text{surr},i}=L_i(W)+\log 2$, and therefore induces the same gradients with respect to $W$. The surrogate is introduced as a practical training objective, not as an exact pointwise reparameterisation of the likelihood.

The surrogate objective also gives a direct update signal for $\sigma_i$. Treating $L_i(W)$ as the BCE term in \eqref{eq:standard-bce}, its derivative with respect to $\sigma_i$ is
\begin{equation}
\setlength{\abovedisplayskip}{2pt}
\setlength{\belowdisplayskip}{2pt}
\frac{\partial \mathcal{L}_{\text{surr},i}}{\partial \sigma_i}
=
-\frac{2L_i(W)}{\sigma_i^3}
+
\frac{1}{\sigma_i + 1}
\label{eq:dsigma}
\end{equation}
This expression clarifies the adaptive nature of CSU. When the class-wise loss remains large relative to the regularisation term, gradient descent increases $\sigma_i$, which in turn reduces the effective coefficient $1/\sigma_i^2$ and weakens the influence of that class on further optimisation updates. 
The logarithmic penalty discourages the trivial solution $\sigma_i \to \infty$. CSU also differs from static class reweighting: a fixed class weight rescales the loss without responding to the class-wise loss observed during training, whereas CSU updates $\sigma_i$ jointly with the network parameters and adapts the effective coefficient $1/\sigma_i^2$ during training.
This adaptive behaviour is also examined in Section~\ref{section_experiments} through the learned $\sigma$ trajectories and the corresponding effective coefficients $1/\sigma_i^2$.

\subsubsection{Final Training Objective and Empirical Expectations}

For a single audio sample, the surrogate losses are summed across classes. Dataset-level training minimises the average of this per-sample objective with respect to the network parameters $W$ and the class-wise supervision unreliability variables $\boldsymbol{\sigma}$. For simplicity, the per-sample objective is written as
\begin{equation}
\setlength{\abovedisplayskip}{2pt}
\setlength{\belowdisplayskip}{2pt}
L(W,\boldsymbol\sigma)
= \sum_{i=1}^{C} \left(
\frac{1}{\sigma_i^2} L_i(W) + \log(\sigma_i + 1)
\right)
\label{final-loss}
\end{equation} 
where $L_i(W)$ denotes the BCE defined in \eqref{eq:standard-bce}. 
This objective makes the effect of CSU explicit in training: classes with less reliable supervision receive smaller effective coefficients $1/\sigma_i^2$ and therefore contribute less to parameter updates, while inference remains unchanged.
If a class is more strongly affected by SAN, MAN, or SLN, it should tend to learn a larger $\sigma_i$ and therefore stronger attenuation through $1/\sigma_i^2$. Section~\ref{section_experiments} therefore examines not only aggregate robustness under SAN, MAN, and SLN, but also the learned $\sigma$ patterns, the effective coefficients $1/\sigma_i^2$, and the associated score-space and local-geometry changes induced by class-wise supervision control.

\vspace{-4mm}
\section{Experiments and Results}
\label{section_experiments}

\vspace{-2mm}
\subsection{Experimental Setup}
Experiments are conducted in two complementary settings. The proposed ESC-FreeGen50 is used for controlled evaluation under SAN, MAN, and SLN. AudioSet \cite{AudioSet} is used for large-scale validation under real-world weakly supervised conditions. ESC-FreeGen50 enables direct analysis of robustness and class-wise supervision control. AudioSet tests whether the same training strategy remains effective beyond the controlled benchmark setting.
  
\noindent\textbf{Dataset:} ESC-FreeGen50 serves as a clean, class-balanced benchmark for robustness evaluation under class-wise supervision unreliability. It contains 5,000 five-second audio clips (6.95 hours), with partitioning following Section~\ref{partitioning}. Its balanced real-and-generated composition and human-verified labels support controlled SAN, MAN, and SLN injection without confounding class-level label quality. 
AudioSet is used for large-scale validation, as described in Section~\ref{audioset_results}.

\noindent\textbf{Baseline:} Google CNN \cite{AudioSet} is used as the primary baseline because it is a canonical and widely used reference model for AT tasks. 
To test whether CSU is architecture-agnostic, the evaluation also includes a representative set of audio tagging backbones covering lightweight CNNs, standard CNNs, large-scale pretrained CNNs, and Transformer-based models, namely MobileNet \cite{sandler2018mobilenetv2}, ResNet \cite{he2016deep}, PANNs \cite{kong2020panns}, and Efficient Audio Transformer (EAT) \cite{chen2024eat}. Because CSU operates at the training objective, it is applied to all backbones without changing network structure or inference.

\noindent\textbf{Corruption Injection Settings:} Evaluation on ESC-FreeGen50 is conducted under controlled SAN, MAN, and SLN injection into the training set, with corruption ratios from 0\% to 50\% in 10\% increments. Corruption is injected independently within each class by uniformly sampling a fixed proportion of clips and applying the corresponding corruption rule. This design enforces class-balanced corruption and supports fair comparison across corruption types and architectures. All corrupted annotations are released on the project homepage (\textcolor{blue}{\underline{https://github.com/Yuanbo2020/CSU}}).
  
\noindent\textbf{Implementation Details:} For ESC-FreeGen50, audio features are 64-bank log-mel energies \cite{logmel}, extracted with a 64 ms Hamming window and 10 ms hop. To ensure fair comparison across architectures, all models use the same input and are trained with Adam \cite{adam}, a learning rate of 0.001, and a batch size of 64. Dropout, normalisation, and early stopping are applied throughout training \cite{dropout}. Training stops when validation performance does not improve for 10 epochs after epoch 20, with a maximum of 100 epochs. AudioSet experiments serve a different role and are reported separately in Section~\ref{audioset_results}. Following standard AudioSet evaluation \cite{kong2020panns}, the unbalanced training set (AS-2M) contains 1,912,134 clips, the balanced validation set (AS-20K) contains 20,550 clips, and the test set contains 18,884 clips across 527 classes.
 
\noindent\textbf{Performance Metrics:} Performance is evaluated using mean Average Precision (mAP), F1-score, Area Under the ROC Curve (AUC), Area Under the Precision–Recall Curve (AUPRC), and exact-match accuracy (Acc) \cite{kong2020panns}.  
Each experiment is repeated 10 times with different random seeds, and the mean and standard deviation are reported.

\vspace{-2mm}
\subsection{Results and Analysis} 

This section evaluates CSU through five research questions (RQ). The analysis moves from controlled validation to broader generalisation. RQ1 establishes the controlled mechanism-level reference on a canonical CNN baseline. RQ2 tests whether the same pattern holds across architectures. RQ3 examines the learned class-wise supervision unreliability parameter $\sigma$ in more detail. RQ4 compares CSU with representative robust-learning methods under matched conditions. RQ5 evaluates whether the advantage of CSU transfers to large-scale real-world weak supervision on AudioSet.

\begin{figure*}[t]
\setlength{\abovecaptionskip}{0cm}   
	\setlength{\belowcaptionskip}{-0.3cm}  
    \centering
    \includegraphics[width=1\linewidth]{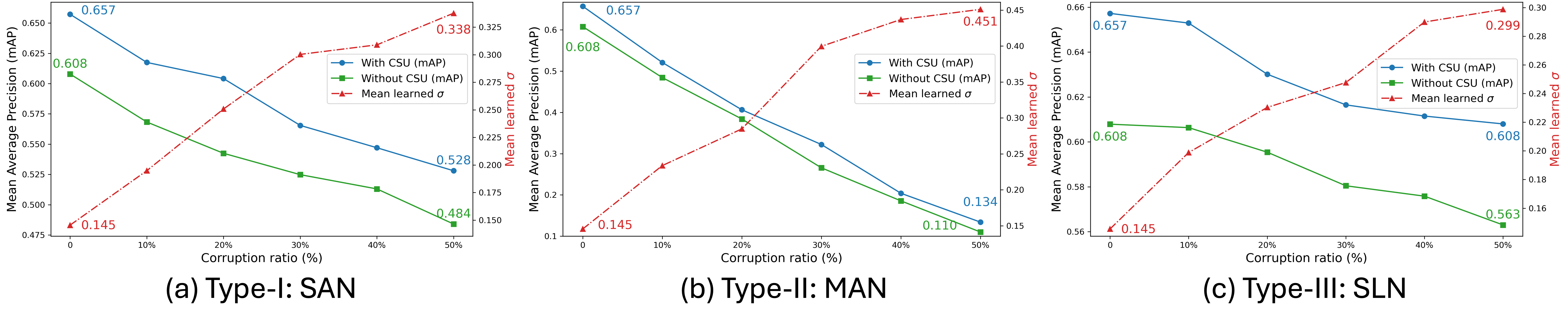}
    \caption{
    Baseline performance under three supervision unreliability types. Each subplot shows mAP for the baseline and the baseline equipped with CSU across corruption ratios from 0\% to 50\% under SAN, MAN, and SLN. The secondary axis shows the mean learned $\sigma$ over the 50 sound classes.
    }
\label{fig:baseline_model}
\end{figure*}

\vspace{-2mm}
\subsubsection{RQ1: Robustness and Response to Different Supervision Corruptions}
  
RQ1 establishes the controlled reference for the rest of this section. Fig.~\ref{fig:baseline_model} shows that CSU improves robustness under SAN, MAN, and SLN on a canonical CNN baseline, while the learned $\sigma$ responds in a mechanism-dependent way.   

\textbf{Type-I: SAN.} 
SAN introduces spurious positive labels while preserving the original class assignment \cite{frenay2013classification}. As shown in Fig.~\ref{fig:baseline_model}(a), increasing SAN causes a gradual decline in mAP for both models. The decrease is moderate because SAN biases supervision without directly corrupting class identity. Across all corruption ratios, CSU remains above the baseline. The learned $\sigma$ also increases steadily under SAN. This trend reveals that accumulated spurious positives make class-wise supervision less reliable. 
Under standard BCE training, such corruption reinforces false activations \cite{natarajan2013learning}. CSU reduces the optimisation impact of these biased updates by weakening the contribution of classes that become unreliable under SAN.
  
\textbf{Type-II: MAN.}
MAN reassigns labels across classes and directly corrupts class identity, causing feature-label contradictions rather than simply adding spurious activations \cite{frenay2013classification}. Fig.~\ref{fig:baseline_model}(b) shows that MAN causes the largest mAP degradation among the three mechanisms. Under standard BCE training, such corrupted labels bias optimisation towards incorrect class associations \cite{natarajan2013learning}. CSU mitigates this effect by reducing the impact of persistently unreliable supervision at the class level. The learned $\sigma$ also shows the largest increase under MAN, indicating that MAN produces the strongest class-wise supervision unreliability among the three controlled settings.
 
\textbf{Type-III: SLN.}
SLN retains the original class assignment but weakens the effective supervision signal by reducing label evidence, thereby simulating ambiguous or low-confidence annotations rather than mislabeling \cite{song2022learning}. In Fig.~\ref{fig:baseline_model}(c), SLN causes the mildest performance degradation. Its main effect is weaker supervision, not structural class corruption. The corresponding increase in $\sigma$ is smooth and limited, suggesting that SLN mainly weakens supervision instead of creating direct class contradiction. In this case, CSU stabilises training under reduced label evidence by moderating class influence without strongly suppressing otherwise correct supervision. 
 
\textbf{Cross-mechanism summary.} 
RQ1 reveals a clear mechanism-dependent ordering. MAN causes the largest drop in mAP and the largest increase in $\sigma$. SAN causes moderate degradation with steadily increasing $\sigma$. SLN causes the mildest degradation and the smallest increase in $\sigma$. Across all three settings, CSU improves performance relative to the baseline.  
The learned $\sigma$ increases with the severity of class-wise supervision unreliability, supporting CSU as a class-wise supervision control mechanism.

\subsubsection{RQ2: Architecture-level Robustness under Supervision Unreliability}

\begin{figure}[t]
\setlength{\abovecaptionskip}{0cm}   
	\setlength{\belowcaptionskip}{-0.1cm}  
    \centering
    \includegraphics[width=1\linewidth]{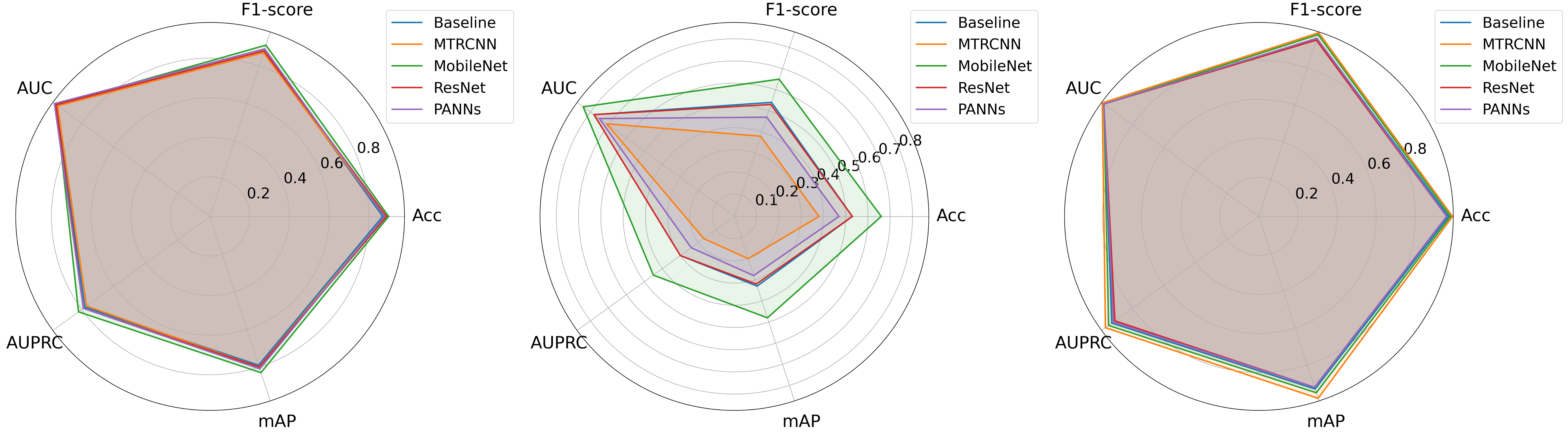} 
    \caption{
    Relative performance change of models under 50\% corruption, normalised to the clean training setting. From left to right: SAN, MAN, and SLN. Lower values indicate stronger degradation.
    }
    \label{fig:robustness}
\end{figure}

RQ2 tests whether the controlled reference in RQ1 remains visible across architectures. Tables~\ref{tab:models_SAN_MAN_SLN_mAP} and~\ref{tab:models_SAN_MAN_SLN_Acc_F1} report mean performance under clean and high-corruption training conditions. Fig.~\ref{fig:robustness} shows the relative degradation ratios normalised to the clean training setting. Across architectures, the same mechanism-dependent ordering remains: MAN causes the strongest degradation, SAN causes intermediate degradation, and SLN causes the mildest degradation. Prior studies also show that label-changing corruption is typically more damaging than weaker or more ambiguous supervision \cite{frenay2013classification, patrini2017making}. Related work on noisy and soft supervision also supports the milder effect of SLN-like conditions when class identity is retained \cite{natarajan2013learning, song2022learning}.

\begin{table}[b]
\footnotesize
\setlength{\tabcolsep}{0pt} 
\setlength{\abovecaptionskip}{0.1cm}   
\setlength{\belowcaptionskip}{-0.7cm}  
\renewcommand{\arraystretch}{0.92} 
\centering
\resizebox{\textwidth}{!}{
\begin{tabular}{
    C{2cm}
    C{2cm}
    C{1cm}C{1cm}C{1cm}C{1cm}
    C{1cm}C{1cm}C{1cm}C{1cm} 
}
\toprule
\multirow{3}{*}{Model} & \multirow{3}{*}{Variants}
& \multicolumn{4}{c}{\makecell[c]{AUPRC (PR-AUC)}}
& \multicolumn{4}{c}{mAP} \\
\cmidrule(lr){3-6}\cmidrule(lr){7-10}
& &
\multirow{2}{*}{0\%} & \multicolumn{3}{c}{corruption ratio 50\%}
& \multirow{2}{*}{0\%} & \multicolumn{3}{c}{corruption ratio 50\%}  \\
\cmidrule{4-6}\cmidrule{8-10} 
& &
& SAN & MAN & SLN
& & SAN & MAN & SLN  \\
\midrule
  
\multirow{2}{*}{Baseline \cite{AudioSet}}  & Base
& 0.573 & 0.453 & 0.085 & 0.543
& 0.608 & 0.484 & 0.110 & 0.563 \\
   & with CSU
& 0.644 & 0.495 & 0.107 & 0.597
& 0.657 & 0.528 & 0.134 & 0.614 \\
\midrule

\multirow{2}{*}{MTRCNN \cite{mtrcnn}}   & Base
& 0.659 & 0.547 & 0.279 & 0.644
& 0.679 & 0.574 & 0.315 & 0.657 \\
   & with CSU
& 0.673 & 0.552 & 0.302 & 0.655
& 0.690 & 0.576 & 0.336 & 0.675 \\
\midrule

\multirow{2}{*}{MobileNet \cite{sandler2018mobilenetv2}}  & Base
& 0.639 & 0.509 & 0.144 & 0.583
& 0.663 & 0.538 & 0.170 & 0.603 \\
       & with CSU
& 0.674 & 0.534 & 0.161 & 0.643
& 0.694 & 0.563 & 0.191 & 0.659 \\
\midrule

\multirow{2}{*}{ResNet \cite{he2016deep}}  & Base
& 0.732 & 0.575 & 0.208 & 0.663
& 0.748 & 0.599 & 0.234 & 0.681 \\
      & with CSU
& 0.742 & 0.588 & 0.221 & 0.678
& 0.757 & 0.607 & 0.241 & 0.694 \\
\midrule

\multirow{2}{*}{PANNs \cite{kong2020panns}}    & Base
& 0.741 & 0.579 & 0.123 & 0.676
& 0.745 & 0.600 & 0.160 & 0.679 \\
        & with CSU
& 0.748 & 0.583 & 0.225 & 0.689
& 0.765 & 0.606 & 0.252 & 0.707 \\
\bottomrule
\end{tabular}
} 
\caption{Mean performance over 10 runs under different supervision unreliability types on the test set (Part 1). Corruption is injected into the training set at 0\% and 50\%, while validation and test labels remain clean. Standard deviations are omitted for brevity.}
\label{tab:models_SAN_MAN_SLN_mAP} 
\end{table}

While preserving this pattern, CSU improves robustness across model families. As shown in Tables~\ref{tab:models_SAN_MAN_SLN_mAP} and~\ref{tab:models_SAN_MAN_SLN_Acc_F1}, models equipped with CSU outperform their corresponding baselines under SAN, MAN, and SLN. Fig.~\ref{fig:robustness} presents the same trend: architecture changes the absolute performance level but does not change the overall corruption ordering or the direction of the CSU gains. Repeated 10-run experiments also support this pattern. Under 50\% MAN, paired t-tests show that the baseline with CSU achieves significantly higher AUPRC and mAP than that without CSU, reaching AUPRC 0.107 and mAP 0.134 (both $p<0.001$). Under the same condition, paired t-tests indicate that MTRCNN with CSU also significantly improves AUPRC and mAP compared to that without CSU, achieving AUPRC 0.302 and mAP 0.336 ($p<0.01$ and $p<0.005$, respectively). Under 50\% SAN, paired t-tests show that MobileNet with CSU achieves significantly higher AUPRC and mAP than that without CSU, reaching AUPRC 0.534 and mAP 0.563 (both $p<0.001$). Similarly, under 50\% MAN, the Wilcoxon signed-rank test shows that MobileNet with CSU also achieves significant improvements in Acc and AUC compared to that without CSU (both $p<0.005$). Under 50\% SLN, paired t-tests show that the baseline with CSU also significantly improves AUPRC and mAP compared to the version without CSU, reaching AUPRC 0.597 and mAP 0.614 (both $p<0.001$). These results show that the advantage of CSU is repeatable across different architectures and corruption settings. 

\begin{table}[b] 
\footnotesize
\setlength{\tabcolsep}{0pt} 
\setlength{\abovecaptionskip}{0.1cm}   
\setlength{\belowcaptionskip}{-0.6cm}  
\renewcommand{\arraystretch}{1} 
\centering
\resizebox{\textwidth}{!}{%
\begin{tabular}{
    C{2cm}
    C{2cm}
    C{1cm}C{1cm}C{1cm}C{1cm}
    C{1cm}C{1cm}C{1cm}C{1cm} 
}
\toprule
\multirow{3}{*}{Model} & \multirow{3}{*}{Variants}
& \multicolumn{4}{c}{\makecell[c]{Acc (\%)}}
& \multicolumn{4}{c}{F1-score} \\
\cmidrule(lr){3-6}\cmidrule(lr){7-10}
& &
\multirow{2}{*}{0\%} & \multicolumn{3}{c}{corruption ratio 50\%}
& \multirow{2}{*}{0\%} & \multicolumn{3}{c}{corruption ratio 50\%}  \\
\cmidrule{4-6}\cmidrule{8-10} 
& &
& SAN & MAN & SLN
& & SAN & MAN & SLN  \\
\midrule

\multirow{2}{*}{Baseline \cite{AudioSet}} & Base
& 75.20 & 66.65 & 26.60 & 73.35
& 0.754 & 0.671 & 0.266 & 0.734 \\
& with CSU
& 79.65 & 69.85 & 30.35 & 76.95
& 0.800 & 0.699 & 0.305 & 0.769 \\
\midrule

\multirow{2}{*}{MTRCNN \cite{mtrcnn}} & Base
& 80.90 & 72.60 & 52.10 & 79.95
& 0.809 & 0.730 & 0.519 & 0.799  \\
& with CSU
& 81.65 & 73.80 & 53.60 & 80.60
& 0.818 & 0.741 & 0.535 & 0.811  \\
\midrule

\multirow{2}{*}{MobileNet \cite{sandler2018mobilenetv2}} & Base
& 79.65 & 70.75 & 36.00 & 76.00
& 0.797 & 0.709 & 0.362 & 0.760 \\
& with CSU
& 81.85 & 72.20 & 38.25 & 79.90
& 0.819 & 0.725 & 0.386 & 0.799  \\
\midrule

\multirow{2}{*}{ResNet \cite{he2016deep}} & Base
& 85.02 & 75.45 & 44.25 & 81.20
& 0.846 & 0.753 & 0.443 & 0.801 \\
& with CSU
& 85.84 & 76.80 & 45.65 & 82.10
& 0.860 & 0.761 & 0.458 & 0.821  \\
\midrule

\multirow{2}{*}{PANNs \cite{kong2020panns}} & Base
& 85.55 & 73.92 & 32.50 & 81.56
& 0.845 & 0.749 & 0.325 & 0.811  \\
& with CSU
& 86.45 & 74.99 & 46.10 & 82.75
& 0.860 & 0.759 & 0.462 & 0.826  \\
\bottomrule
\end{tabular}%
}
\caption{Mean test performance over 10 runs under different supervision unreliability types (Part 2).}
\label{tab:models_SAN_MAN_SLN_Acc_F1}
\end{table}

Two secondary observations emerge. First, ranking-based metrics (e.g., AUC) are more stable than precision-recall metrics under corruption, as shown in Fig.~\ref{fig:robustness}. AUPRC and mAP degrade more sharply, which indicates that corrupted supervision affects confidence quality more strongly than coarse ranking quality. Similar behaviour has also been reported in noisy audio tagging and weakly labeled audio datasets \cite{fonseca2019learning, zhu2020audio}. Second, architecture changes the magnitude of the degradation, but not its overall direction. Models with stronger clean-label performance can still show substantial drops under severe corruption, while the same corruption ordering and the same direction of CSU gains remain visible across architectures. 
 
SLN remains the mildest condition across architectures. Because SLN weakens label evidence without changing class identity, it introduces less structurally unreliable supervision than SAN or MAN. Under this condition, CSU acts more as mild supervision control than as protection against structural corruption. Overall, RQ2 extends the controlled reference from RQ1 to a broader set of backbones: corruption structure determines the overall pattern, and CSU improves robustness across architectures without changing that pattern.  

 
\begin{figure*}[t]
\setlength{\abovecaptionskip}{0cm}   
	\setlength{\belowcaptionskip}{-0.1cm}  
    \centering
    \includegraphics[width=1\linewidth]{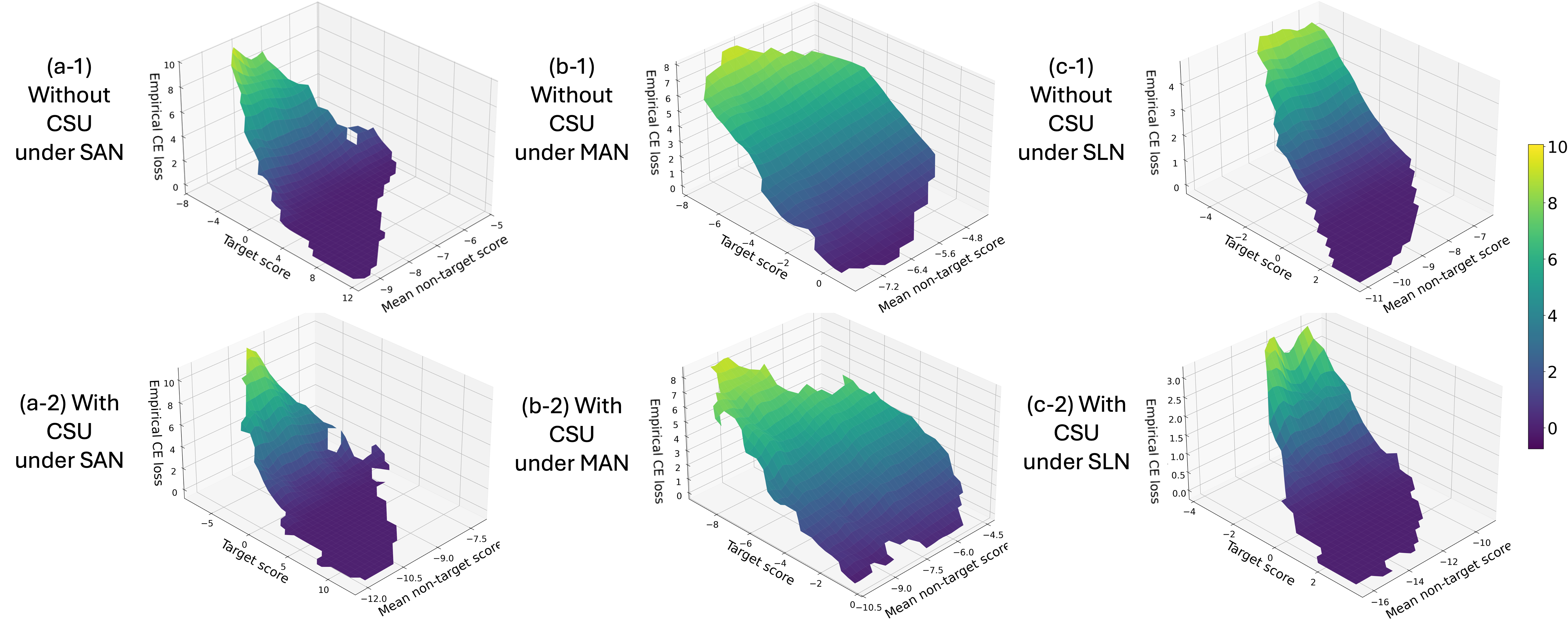} 
    \caption{Score-plane loss surfaces under three 50\% corruption settings. Each sample is mapped to a plane defined by the target-class score and the mean non-target score, and the surface height shows empirical cross-entropy (CE) loss. Each surface is aggregated over the 10 runs used for the analysis.}
    \label{fig:loss_surfaces}
\end{figure*}

\subsubsection{RQ3: Learned $\sigma$ Patterns under Different Supervision Unreliability Types}
\label{CSU_loss_surface_gradient_multiplier}
  
RQ3 studies whether CSU learns a control signal that changes with the structure of supervision unreliability. The analysis has two linked parts. The first examines the optimisation effects induced by CSU on the baseline under 50\% SAN, MAN, and SLN. The second examines the learned parameter \(\sigma\) itself across architectures.

\begin{figure}[t]
\setlength{\abovecaptionskip}{0.1cm}    
	\setlength{\belowcaptionskip}{-0.1cm}  
    \centering
    \includegraphics[width=0.8\linewidth]{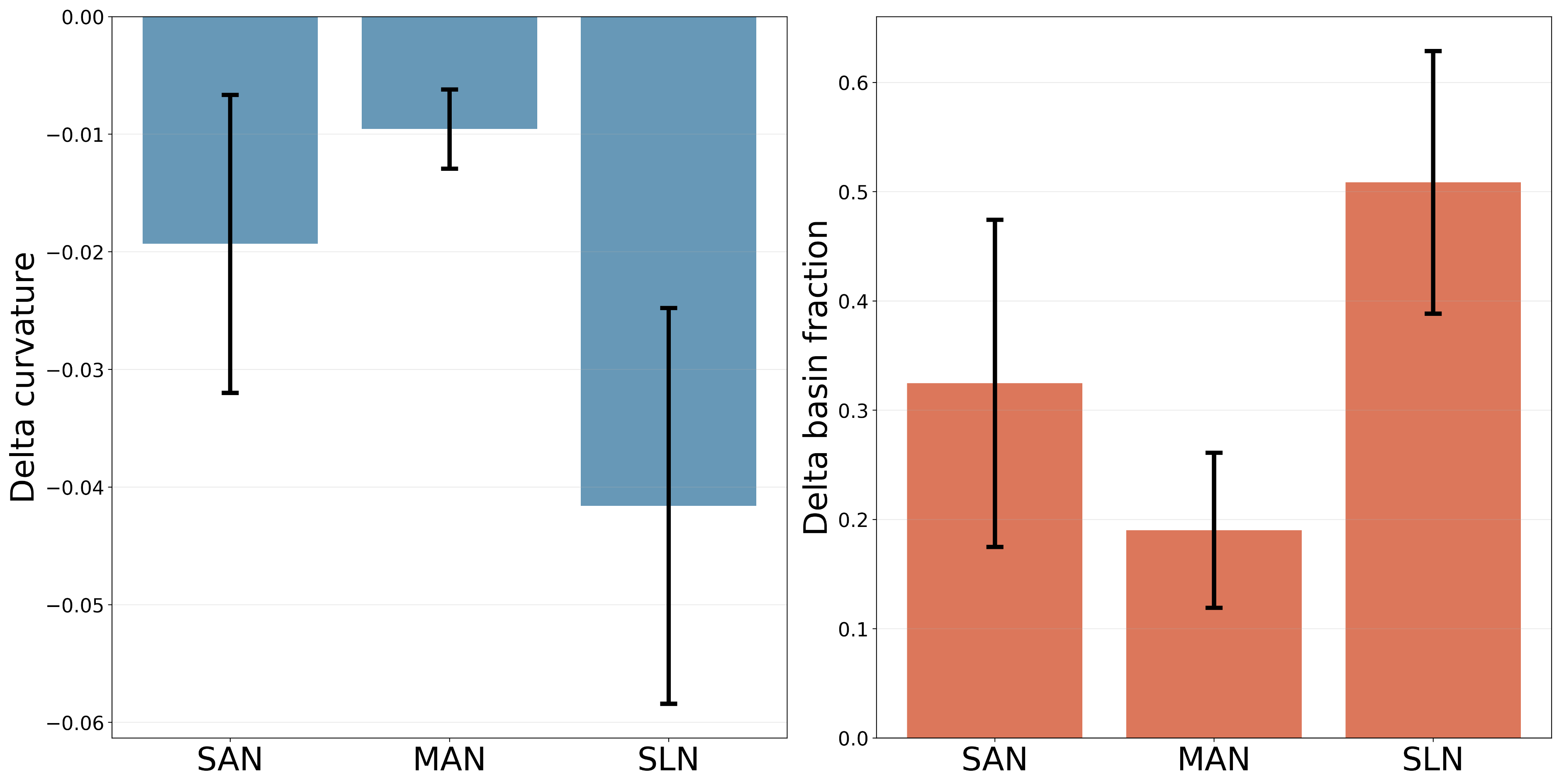} 
    \caption{Summary statistics of local parameter-space geometry based on the loss landscapes in Fig.~\ref{fig:loss_surfaces}, aggregated over the 10 runs for the analysis, with error bars denoting 95\% confidence intervals. The left panel shows curvature difference and the right panel shows the low-loss basin-fraction difference, both defined as $Delta=$ (with CSU) $-$ (without CSU), under 50\% SAN, 50\% MAN, and 50\% SLN. Negative curvature differences and positive basin-fraction differences indicate flatter and wider local geometry.} 
    \label{fig:loss_summary}
\end{figure} 

\begin{figure*}[t]
\setlength{\abovecaptionskip}{0.1cm}    
	\setlength{\belowcaptionskip}{-0.3cm}  
    \centering
    \includegraphics[width=1\linewidth]{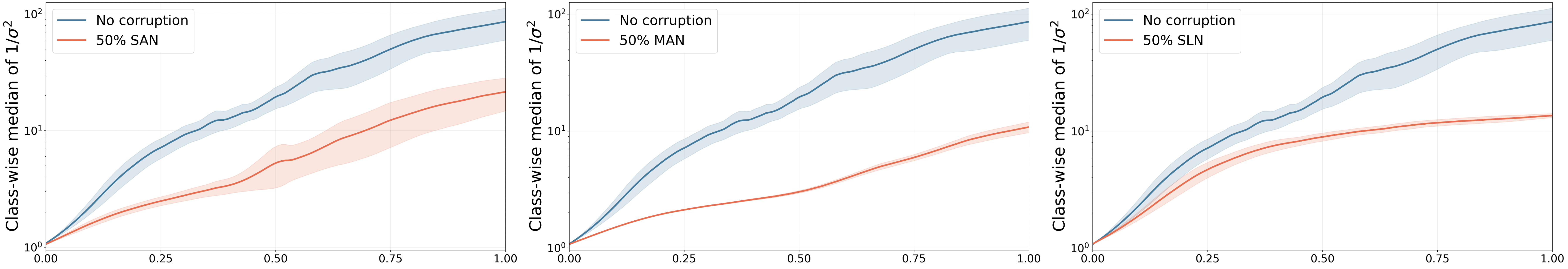} 
    \caption{Effective gradient multiplier trajectories learned by CSU. The x-axis denotes normalised training progress, and the y-axis denotes the class-wise median CSU multiplier \(1/\sigma^2\) (log scale).  The trajectories are averaged over the 10 CSU runs for each condition under 0\% corruption, 50\% SAN, 50\% MAN, and 50\% SLN, with shaded bands denoting 95\% confidence intervals.}
    \label{fig:sigma_CSU}
\end{figure*}

The first step analyses three connected views of optimisation: score space, local parameter-space geometry, and the effective coefficient \(1/\sigma^2\) that scales each class-wise loss term and gradient. Fig.~\ref{fig:loss_surfaces} shows how CSU reshapes loss pressure in score space. Under SAN, MAN, and SLN, the baseline with CSU shows more compact high-loss structures and a larger accessible low-loss region than the baseline without CSU. 
Fig.~\ref{fig:loss_summary} shows the same pattern in local parameter-space geometry. With CSU, the summary statistics show flatter and wider local geometry around the trained model, reflected by lower local curvature and a larger low-loss basin.
Fig.~\ref{fig:sigma_CSU} shows the class-wise median trajectories of \(1/\sigma^2\). In CSU, \(1/\sigma^2\) is the effective coefficient on each class-wise loss term and its corresponding gradient, not the final weighted loss value. Higher values indicate stronger effective class-wise updates, whereas lower values indicate stronger suppression. Under 50\% corruption, these coefficients remain below those of no corruption throughout training, which means that CSU persistently weakens the update strength assigned to corrupted supervision.
  
These three views describe the same mechanism at different levels. Lower effective coefficients \(1/\sigma^2\) reduce the influence of unreliable supervision during optimisation. The resulting loss structure becomes less dominated by high-loss regions in score space, and the local geometry of the model becomes flatter and wider. The strength of this mechanism differs across corruption types. Under SAN, CSU suppresses excess supervision pressure introduced by spurious positives. Under MAN, CSU reduces the optimisation damage caused by contradictory supervision between confusable classes. Under SLN, CSU mainly recalibrates update strength when label evidence is weakened, but class identity is preserved.

\begin{figure}[h]
\setlength{\abovecaptionskip}{0cm}   
	\setlength{\belowcaptionskip}{-0.4cm}  
    \centering
    \includegraphics[width=1\linewidth]{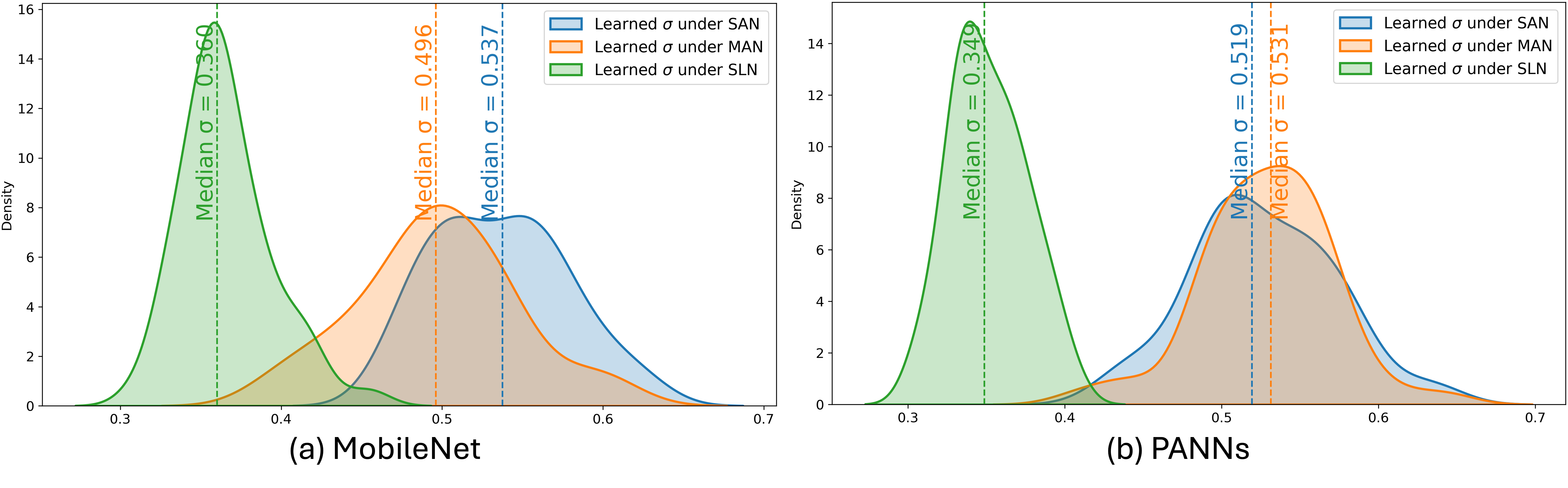} 
\caption{Kernel density estimates of the learned $\sigma$ under SAN, MAN, and SLN at a 50\% corruption ratio for two representative backbones: (a) MobileNet and (b) PANNs. Each curve represents the distribution of $\sigma$ values across the 50 sound event classes.}
\label{fig:KED_MobileNet_PANN} 
\end{figure}

The second step turns to the learned parameter \(\sigma\) itself. If the control mechanism proposed in Section~\ref{section_clu} is valid, then the learned \(\sigma\) should vary with supervision mechanism, not only with the backbone. To test this, MobileNet \cite{sandler2018mobilenetv2} and PANNs \cite{kong2020panns} are used as representative audio tagging backbones, covering a lightweight model and a stronger pretrained model. Fig.~\ref{fig:KED_MobileNet_PANN} shows kernel density estimates of \(\sigma\) under SAN, MAN, and SLN at the 50\% corruption ratio reported in Table~\ref{tab:models_SAN_MAN_SLN_mAP}. Across both models, one contrast remains stable. SLN produces the lowest and most concentrated \(\sigma\) distribution. SAN and MAN produce broader distributions with higher overall levels. The relative separation between SAN and MAN depends on the backbone, but the main contrast between SLN and the other two corruption types remains clear.

\textbf{SAN: broader and more dispersed \(\sigma\) under spurious positives.}
Under SAN, Fig.~\ref{fig:KED_MobileNet_PANN} shows broad \(\sigma\) distributions rather than compact single peaks. That is, SAN does not weaken supervision uniformly across classes. Some classes accumulate more spurious positives than others, so the learned \(\sigma\) spreads over a wider range. Weakly labeled corpora such as AudioSet \cite{AudioSet} and FSDnoisy18k \cite{fonseca2019learning} also show uneven exposure to spurious contamination. The broader spread of \(\sigma\) under SAN shows that CSU adjusts class influence unevenly when spurious positives accumulate.

\textbf{MAN: increased \(\sigma\) under contradictory supervision.}
Under MAN, Fig.~\ref{fig:KED_MobileNet_PANN} shows \(\sigma\) distributions that shift upward relative to SLN and remain broad across the label set. MAN directly reassigns labels across classes and introduces contradictory feature-label associations throughout training \cite{natarajan2013learning, patrini2017making}. This broad contradiction lifts the \(\sigma\) distribution as a whole. 
The higher overall \(\sigma\) distributions under MAN show that CSU applies stronger class-wise suppression in this setting.

\textbf{SLN: low and concentrated \(\sigma\) under weakened label evidence.}
Under SLN, Fig.~\ref{fig:KED_MobileNet_PANN} shows the lowest and narrowest \(\sigma\) distributions. SLN weakens supervision evidence without creating strong class contradiction. Class identity is preserved, and the reduction in supervision strength is more uniform across classes. The learned \(\sigma\) remains relatively low and concentrated under SLN, instead of showing the broader distributions observed under SAN and MAN.

\textbf{Cross-architecture summary.}  
RQ3 links the optimisation analysis on the baseline to the learned \(\sigma\) distributions. On the baseline, CSU reduces the effective coefficient \(1/\sigma^2\), contracts high-loss structures, and flattens local geometry. 
Across architectures, the learned \(\sigma\) also varies with the supervision mechanism. SLN remains low and concentrated in both models, whereas SAN and MAN produce broader and higher distributions. The relative separation between SAN and MAN depends on the backbone, but the stable contrast between SLN and the other two corruption types remains.
These results show that CSU learns a class-wise control signal whose parameter pattern and optimisation effect both vary with the structure of supervision unreliability.

\subsubsection{RQ4: Comparison with Representative Robust-Learning Methods}

RQ1--RQ3 investigate how CSU behaves under controlled SAN, MAN, and SLN, how the same pattern generalises to architectures, and how the learned control signal changes with the supervision mechanism. RQ4 then compares CSU with representative robust learning methods under the same controlled setting with 50\% mixed corruption (SAN/MAN/SLN = 1:1:1). The equal mixture is used because large-scale weakly labeled audio corpora such as AudioSet \cite{AudioSet} and FSD50K \cite{Fonseca2022fsd50k} often contain spurious positives, class confusions, and weakened label evidence together rather than in isolation. The 50\% ratio keeps the setting difficult enough to distinguish methods while remaining within a reasonable range of weak supervision.  

As described in Section~\ref{Introduction}, ASL \cite{ridnik2021_asl} and APL \cite{huang2023apl} represent gradient-shaping methods for weak multi-label learning, SCE \cite{wang2019sce} represents robust loss design, Bootstrapping \cite{Bootstrapping} represents soft-target correction, UPS \cite{ups} represents uncertainty-filtered pseudo-label selection, and $\rho$-corrected DC \cite{lowne2010sequential} represents explicit global corruption correction. These methods cover the main comparison directions relevant to this study. All methods are evaluated on the same MobileNet backbone, which is lightweight, stable, and already used in the controlled analyses above.   
This keeps the comparison focused on method-level differences and avoids mixing robustness effects with architecture-specific capacity.

\begin{table}[t]
\footnotesize
\setlength{\tabcolsep}{4pt} 
\setlength{\abovecaptionskip}{0.1cm}   
\setlength{\belowcaptionskip}{-0.2cm}  
\renewcommand{\arraystretch}{1} 
\centering
\resizebox{\textwidth}{!}{
\begin{tabular}{
    C{1.5cm}
    C{1.8cm}
    C{1.8cm}
    C{1.8cm}
    C{1.8cm}
}
\toprule
$\rho$ & Acc (\%) & F1 & PR-AUC & mAP \\
\midrule
0     & $60.78 \pm 1.94$ & $0.610 \pm 0.020$ & $0.380 \pm 0.024$ & $0.412 \pm 0.024$ \\
0.005 & $61.92 \pm 2.10$ & $0.622 \pm 0.020$ & $0.394 \pm 0.025$ & $0.429 \pm 0.023$ \\
0.01  & $63.40 \pm 1.18$ & $0.636 \pm 0.010$ & $0.412 \pm 0.013$ & $0.441 \pm 0.013$ \\
0.025 & $64.12 \pm 0.41$ & $0.644 \pm 0.001$ & $0.421 \pm 0.001$ & $0.453 \pm 0.008$ \\
0.03  & $61.16 \pm 2.39$ & $0.612 \pm 0.023$ & $0.382 \pm 0.027$ & $0.416 \pm 0.027$ \\
\bottomrule
\end{tabular}
}
\caption{Test-set performance of $\rho$-corrected DC \cite{lowne2010sequential} across $\rho$ under 50\% mixed label corruption (SAN/MAN/SLN = 1:1:1).}
\label{tab:rt_dc_results}
\end{table}

\textbf{Performance of $\rho$-corrected DC under different $\rho$.} 
Among the compared methods, $\rho$-corrected DC \cite{lowne2010sequential} requires explicit selection of its core parameter $\rho$, which is part of the method definition and directly controls the strength of probability correction under the bit-flip model. To illustrate this, Table~\ref{tab:rt_dc_results} reports the performance of $\rho$-corrected DC across different $\rho$ values with the same validation-based early-stopping strategy. Performance improves from 0 to 0.025 on all reported metrics, then drops at 0.03, suggesting that mild correction helps under mixed corruption, whereas stronger correction begins to suppress useful supervision along with corrupted supervision.  
 
\textbf{Comparison with other methods.} Table~\ref{tab:robust_mobilenet_50mix} reports the main comparison. SCE performs the worst on all reported metrics. ASL performs better than SCE but remains below the stronger methods. APL, Bootstrapping, and UPS form a middle group with similar results. CSU and $\rho$-corrected DC outperform the remaining methods under this setting. The ordering reflects the different assumptions of each method. 
ASL \cite{ridnik2021_asl} and APL \cite{huang2023apl} mainly reshape gradients to handle multi-label imbalance and easy-negative dominance, but they do not explicitly model class-wise supervision reliability. Bootstrapping \cite{Bootstrapping} softens corrupted targets by mixing labels with predictions, but it can still inherit early prediction bias. UPS \cite{ups} filters pseudo-labels with uncertainty estimates, but its gains depend on thresholding and selection rather than direct class-wise supervision control. SCE \cite{wang2019sce} relies on robust loss design, yet in this mixed setting, it appears less suited to the combined presence of spurious positives, class misassignment, and weakened label evidence.

CSU and $\rho$-corrected DC show close performance in Table~\ref{tab:robust_mobilenet_50mix}. $\rho$-corrected DC assumes that corruption can be handled through a global bit-flip correction parameter, whereas CSU learns class-wise supervision control without reducing mixed unreliable supervision to a single global flip-rate. This distinction matters here because SAN, MAN, and SLN affect supervision in different ways and do not act uniformly across classes. Under this condition, CSU remains highly competitive while staying closer to the class-wise structure of mixed supervision unreliability.

In short, RQ4 places CSU in a clearer method-level position. Under mixed corruption, CSU remains competitive among the compared methods because its class-wise control mechanism is better matched to mixed supervision unreliability than methods built around gradient reshaping, target softening, or global correction alone. The next RQ tests whether the same advantage remains on AudioSet.
\begin{table}[t]
\footnotesize
\setlength{\tabcolsep}{1pt} 
\setlength{\abovecaptionskip}{0.1cm}   
\setlength{\belowcaptionskip}{-0.2cm}  
\renewcommand{\arraystretch}{1} 
\centering
\begin{tabular}{
    C{4.7cm}
    C{1.8cm}
    C{1.7cm}
    C{1.8cm}
    C{1.7cm}
}
\toprule
Method & Acc (\%) & F1 & ROC-AUC & mAP \\
\midrule
Asymmetric Loss (ASL) \cite{ridnik2021_asl} & $53.88 \pm 1.05$ & $0.550 \pm 0.006$ & $0.825 \pm 0.006$ & $0.343 \pm 0.006$ \\
Asymmetric Polynomial Loss (APL) \cite{huang2023apl} & $60.72 \pm 2.09$ & $0.611 \pm 0.018$ & $0.807 \pm 0.009$ & $0.409 \pm 0.017$ \\
Symmetric Cross Entropy (SCE) \cite{wang2019sce} & $38.96 \pm 1.13$ & $0.390 \pm 0.011$ & $0.689 \pm 0.006$ & $0.194 \pm 0.012$ \\
Bootstrapping \cite{Bootstrapping} & $61.20 \pm 1.51$ & $0.612 \pm 0.015$ & $0.805 \pm 0.007$ & $0.417 \pm 0.017$ \\
UPS \cite{ups} & $61.60 \pm 0.81$ & $0.615 \pm 0.012$ & $0.808 \pm 0.007$ & $0.420 \pm 0.011$ \\
$\rho$-corrected DC \cite{lowne2010sequential} & $64.12 \pm 0.41$ & $0.644 \pm 0.001$ & $0.820 \pm 0.001$ & $0.453 \pm 0.008$ \\
Proposed CSU & $64.32 \pm 0.88$ & $0.645 \pm 0.006$ & $0.823 \pm 0.003$ & $0.453 \pm 0.010$ \\
\bottomrule
\end{tabular}
\caption{Test-set performance of methods under 50\% mixed label corruption (SAN/MAN/SLN = 1:1:1).}
\label{tab:robust_mobilenet_50mix}
\end{table}
 
\subsubsection{RQ5: Generalisation to Large-Scale Weak Supervision on AudioSet}\label{audioset_results} 
RQ5 moves from controlled corruption to large-scale real-world weak supervision. Unlike RQ1--RQ4, AudioSet \cite{AudioSet} does not allow SAN, MAN, and SLN to be separated and analysed individually. Instead, they coexist within weak clip-level labels. RQ5 tests whether CSU still improves learning under this setting.

\textbf{Models and evaluation setup.}
AudioSet experiments use the Efficient Audio Transformer (EAT) \cite{chen2024eat}, a strong Transformer-based audio tagging model built on AST \cite{ast} with audio self-supervised pretraining. To separate architectural modification from CSU itself, two variants are compared. EAT is the original backbone. EAT-CH replaces the original classification layer with 527 class-specific projection heads of dimension 64, one for each AudioSet class. EAT-CSU builds on EAT-CH and adds CSU to the training objective. This design makes it possible to distinguish gains from class-specific output heads from gains produced by CSU. 

All models follow the EAT recipe \cite{chen2024eat} for data sampling, augmentation, and optimisation settings. To accelerate training, the EAT backbone is loaded and frozen when training EAT-CH, and only the 527 class-specific heads are updated. EAT-CH is trained with a batch size of 256 and a learning rate of 5e-5. Training stops if validation mAP does not improve for 1,000 iterations, with a maximum of 25,000 iterations. EAT-CSU is initialised from the trained EAT-CH and trained with the same batch size, learning rate, and early-stopping criterion, with a maximum of 35,000 iterations. All other settings follow the EAT recipe. For AudioSet \cite{AudioSet}, AS-2M with 1,912,134 clips is used for training, AS-20K with 20,550 clips for validation, and the evaluation set with 18,884 clips for testing. Performance is reported on the evaluation set \cite{AudioSet}.

\begin{table}[t]
\footnotesize
\setlength{\tabcolsep}{1pt} 
\setlength{\abovecaptionskip}{0.2cm}   
\setlength{\belowcaptionskip}{-0.2cm}  
\renewcommand{\arraystretch}{1} 
\centering  
\begin{tabular}{    
C{2cm}
    C{5.7cm}
    C{1.8cm}
    C{2cm} }
\toprule
Category  & Model & \# Parameters & mAP (\%) \\
\midrule
\multirow{4}{*}{\makecell{Supervised\\Learning}} & Pretrained Audio Neural Networks (PANNs)~\cite{kong2020panns} & 81 M & 43.1 \\
  & Audio Spectrogram Transformer (AST)~\cite{ast} & 86 M & 45.9 \\
  & Multimodal Bottleneck Transformer (MBT)~\cite{nagrani2021attention} & 86 M & 44.3 \\
  & AudioCLIP~\cite{guzhov2022audioclip} & 93 M & 25.9 \\
\midrule
\multirow{5}{*}{\makecell{Self-Supervised\\Learning}} & Conformer-based Self-Supervised Learning~\cite{srivastava2022conformer} & 88 M & 41.1 \\
 & Audio Masked Autoencoder (AudioMAE)~\cite{huang2022masked} & 86 M & 47.3 \\
 & BEATs~\cite{chen2023beats} & 90 M & 48.0 \\
 & Masked Spectrogram Prediction (MaskSpec)~\cite{chong2023masked} & 86 M & 47.1 \\
 & Efficient Audio Transformer (EAT)~\cite{chen2024eat} & 88 M & 48.6 \\
\midrule
\multirow{3}{*}{\makecell{Proposed}} & EAT-CH & 111 M &  \textbf{49.02$\pm$0.18} \\
 & \multirow{2}{*}{EAT-CSU} & \multirow{2}{*}{111 M} & \textbf{49.61$\pm$0.27} \\
 & & & (Best 50.04)\\
\bottomrule
\end{tabular}
\caption{Mean Average Precision (mAP) of audio tagging models on the AudioSet evaluation set. Results for the proposed models are reported as mean $\pm$ standard deviation over 10 runs. Results for the other models are taken from the corresponding references.}
\label{tab:audioset_models}
\end{table} 

\textbf{Results on AudioSet.} 
Table~\ref{tab:audioset_models} places the models in the context of recent supervised and self-supervised AudioSet systems. Supervised models remain clearly below the strongest self-supervised models, confirming the importance of large-scale audio pretraining on this benchmark. Within the self-supervised group, EAT is already a strong audio-only reference. EAT-CH further improves over EAT, which means class-specific output heads provide additional flexibility. EAT-CSU improves over EAT-CH, indicating that the gain is not explained by architectural modification alone. With 49.61\% mAP on average and 50.04\% at best on the AudioSet evaluation set, EAT-CSU shows that the benefit of CSU transfers from the controlled benchmark to large-scale real-world weak supervision.

\begin{figure}[H]
\setlength{\abovecaptionskip}{0cm}   
	\setlength{\belowcaptionskip}{-0.4cm}  
    \centering
    \includegraphics[width=1\linewidth]{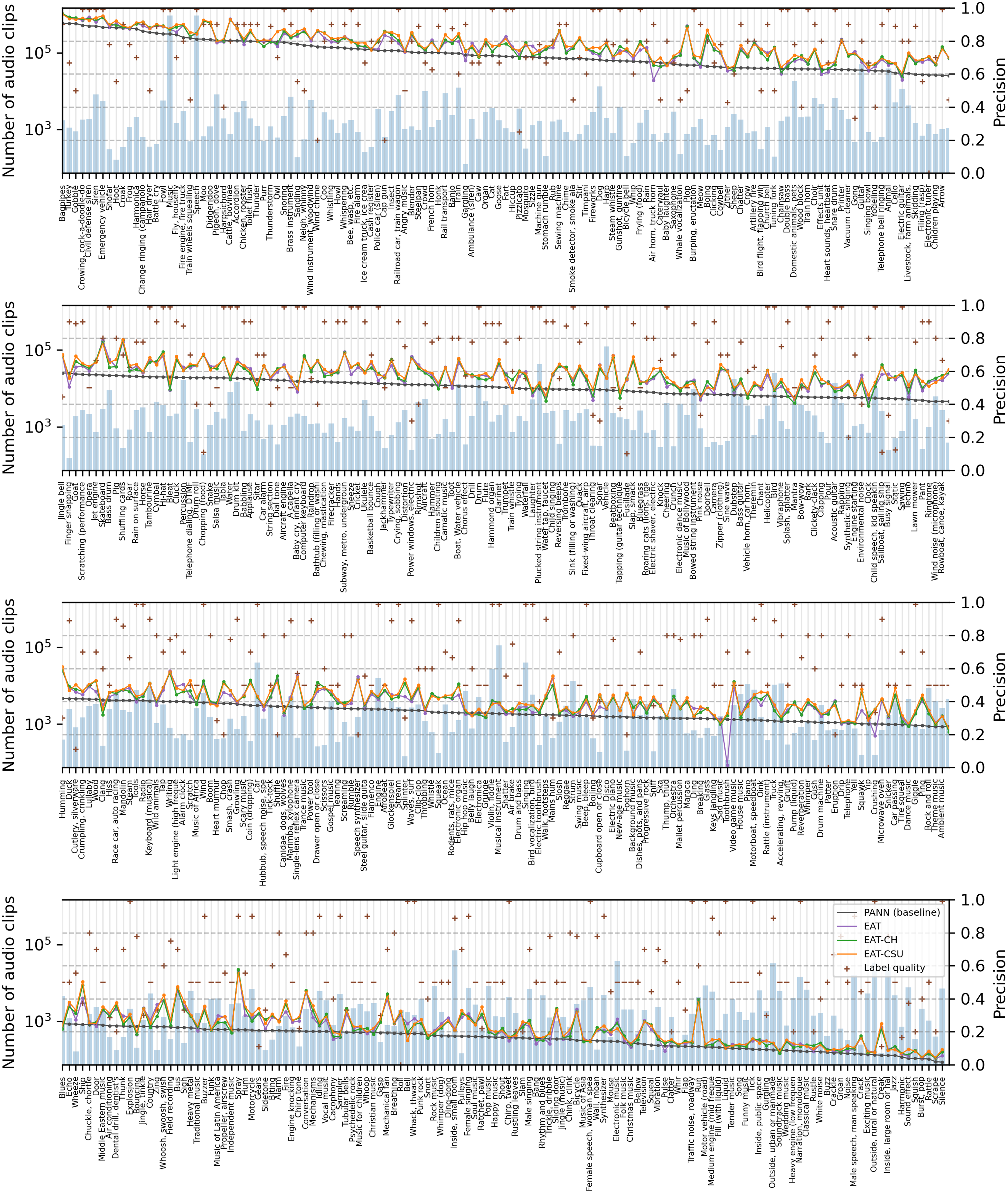}
    \caption{Class-wise precision of audio tagging (AT) models on the AudioSet evaluation set \cite{AudioSet} (527 event classes).
Black, purple, green, and orange curves show the results of PANNs \cite{kong2020panns}, EAT \cite{chen2024eat}, and the proposed EAT-CH and EAT-CSU, respectively. 
Blue bars show the number of training clips on logarithmic scale, and plus markers denote the class-wise label quality provided by AudioSet.
}  
\label{fig:audioset_models_mAP}
\end{figure}

\textbf{Why the gain is reasonable under weak labels.}
AudioSet labels are weak at the clip level, and overlapping events make supervision incomplete and uneven across classes \cite{AudioSet}. Under such conditions, SAN-like effects, MAN-like effects, and weakened label evidence can coexist without being separable during training. CSU works because it does not require these sources to be isolated. It requires supervision reliability to differ across classes strongly enough to affect learning.
 
\textbf{Class-wise precision profile.} 
Fig.~\ref{fig:audioset_models_mAP} compares class-wise precision together with class frequency and class-wise label quality on AudioSet. All models show variations across classes, indicating that weak supervision quality is not uniform. This variation is not explained by class frequency alone, because even frequent classes can show low precision when labels are weak or ambiguous. Compared with PANNs, EAT, and EAT-CH, EAT-CSU shows a more stable class-wise precision profile, especially in low- and mid-frequency classes. The comparison between EAT-CH and EAT-CSU shows that EAT-CH increases class-specific capacity, and EAT-CSU adds a further stabilising effect beyond that change.
  
Overall, RQ5 completes the transition from controlled corruption to real weak supervision. On AudioSet, CSU improves a strong Transformer-based baseline without adding inference complexity. The improvement holds when supervision unreliability is mixed, large-scale, and impossible to isolate. This result shows that CSU remains useful beyond controlled benchmark settings.

\vspace{-2mm}
\section{Conclusion}\label{section_conclusion} 

This paper studies class-wise supervision unreliability in weakly labeled audio tagging and introduces CSU as a training framework for class-wise supervision control. The problem becomes especially important when weak labels are incomplete, contradictory, or uneven across classes, particularly in recent pipelines that mix real and generated audio. To study this problem under controlled conditions, the paper also introduces ESC-FreeGen50, which supports direct analysis of SAN, MAN, and SLN in a clean, label-verified real-and-generated benchmark setting.  

Experiments show that SAN, MAN, and SLN produce different degradation patterns and different learned control responses. MAN causes the strongest degradation, SAN causes intermediate degradation, and SLN causes the mildest degradation. CSU responds to these differences through mechanism-dependent changes in the learned \(\sigma\), the effective coefficients \(1/\sigma^2\), and the resulting optimisation behaviour.  
The same pattern appears across architectures, and CSU remains competitive against representative robust-learning methods under matched mixed corruption. The gain also transfers from ESC-FreeGen50 to AudioSet, which shows that CSU remains effective in both controlled corruption analysis and large-scale real-world weak supervision.   

Overall, the results show that class-level modelling of supervision reliability is a useful alternative to instance-level correction when weak labels do not expose a reliable corruption path. This is particularly relevant to large-scale weakly labeled recognition problems, where supervision is mixed, class-dependent, and difficult to disentangle. In this sense, CSU is useful not only for controlled analysis, but also as a practical training strategy for large-scale weak supervision.

The main strength of CSU is that it provides a simple and scalable way to reduce the optimisation influence of persistently unreliable supervision without changing the network architecture or the inference process. The main limitation is that CSU uses one learned scalar per class, so it cannot capture finer-grained variation across instances or time. Future work will extend class-wise supervision control to finer-grained settings, including sound event detection, streaming audio, and domain shift.

\bibliographystyle{elsarticle-num}
\bibliography{cas-refs}
\end{document}